\documentclass{Interspeech2024}
\usepackage{multirow} 
\usepackage{diagbox}

\newcommand{\asr}{g_{\phi}}
\newcommand{\adaptor}{f_{\theta}}
\newcommand{\WER}{\mathrm{WER}}
\newcommand{\adapter}{adapter}
\newcommand{\ours}{Ours}
\newcommand{\other}{tPLCnet}

\interspeechcameraready

\title{Enhanced ASR Robustness to Packet Loss with a Front-End Adaptation Network}

\name{Yehoshua}{Dissen}
\name{Shiry}{Yonash}
\name{Israel}{Cohen}
\name{Joseph}{Keshet}

\address{
  Faculty of Electrical and Computer Engineering,\\ Technion -- Israel Institute of Technology, Israel}
\email{shua.dissen@gmail.com}


\begin{document}

\maketitle
 

\begin{abstract}
In the realm of automatic speech recognition (ASR), robustness in noisy environments remains a significant challenge. Recent ASR models, such as Whisper, have shown promise, but their efficacy in noisy conditions can be further enhanced. This study is focused on recovering from packet loss to improve the word error rate (WER) of ASR models. We propose using a front-end adaptation network connected to a frozen ASR model. The adaptation network is trained to modify the corrupted input spectrum by minimizing the criteria of the ASR model in addition to an enhancement loss function. Our experiments demonstrate that the adaptation network, trained on Whisper's criteria, notably reduces word error rates across domains and languages in packet-loss scenarios. This improvement is achieved with minimal affect to Whisper model's foundational performance, underscoring our method's practicality and potential in enhancing ASR models in challenging acoustic environments. \end{abstract}

\section{Introduction} \label{intro}
Automatic speech recognition (ASR) has made huge progress in the past few years with the introduction of large pre-trained \cite{baevski2020wav2vec, hsu2021hubert} or weakly supervised transformer models trained on massive amounts of data, such as Whisper \cite{radford2023robust}. While these models perform well in many domains, there still is room for significant improvement in challenging environments such as noisy or reverberant data. This work focuses on one of the more complex cases, namely, the packet loss scenario, where parts of the audio data are lost or corrupted during transmission.  

Improving the models' robustness under these conditions can be challenging. Finetuning these models can easily overfit to the domain of data you finetune with. For example, if the model is finetuned on English data, the model might ``forget'' other languages \cite{sun2023can}. Even within the same language, the model can improve on read speech while degrading performance on phone call speech. Training a model from scratch is often impractical due to the large size of the model and the vast amount of data required, which can demand excessive computational resources and time.

Another approach, would be using a packet loss concealment (PLC) model. PLC algorithms aim to solve the task of reconstructing missing frames from a signal. This work seeks to create a packet loss concealer directed at the downstream task of improving automatic speech recognition (ASR). 
Most PLC algorithms aim to improve the speech quality perception, while very few aim to solve a downstream task such as \cite{mohamed2020concealnet}, where they work on PLC for speech emotion recognition. 
Traditionally, concealment was done using some form of Linear Prediction or interpolation algorithms such as in \cite{gunduzhan2001linear, chen2009packet, ofir2007audio}. Following the proliferation of neural nets, they have become the predominant method \cite{wang2021temporal, westhausen2022tplcnet, pascual2021adversarial, lin2021time}. These models generally improve human intelligibility, and usually improve WER as well; however, they can introduce artifacts or distortions in the signal that are not well-received by an ASR model, compromising their efficacy. We further discuss related work in Section \ref{related}. 

We aim to create a simple method for improving packet loss robustness for foundational ASR models without needing in-domain data. For this, we turned to the ASR model inputs or features. We added a small front-end adaptation model that fills the gaps in the input spectrum before passing it on to the backend ASR model. We used a U-net \cite{ronneberger2015u} architecture with skip connections from popular PLC and inpainting models. However, since our goal is to improve ASR metrics, specifically Word Error Rate (WER) and not the audio quality, instead of using perceptual losses, we utilize the gradients from the ASR model to update the adaptation models weights. We are essentially training a packet loss concealer with ASR objectives while keeping the ASR model frozen.
This strategy allows for improving the model's robustness to packet loss while keeping the ASR model's initial capabilities without retraining the entire model, a process often constrained by resource limitations or fine-tuning where the model is prone to domain overfitting.
A comprehensive series of experiments with the proposed method demonstrated greatly improved robustness to packet-loss corruption compared to the baseline models, fine-tuning, and other PLC methods. This includes performance across domains and languages that differed from those in the training set. Moreover, by maintaining the ASR models' weights unchanged, the original performance was not compromised, ensuring that the improvements in robustness did not detract from the models existing capabilities.

The main contribution of this work is a method that can improve foundational ASR models' robustness to packet loss without changing the underlying ASR model or degrading its results using a very lightweight \adapter{} model. Our implementation and trained models are available here \footnote{{\url{https://github.com/MLSpeech/WhisperDenoiser}}}.


\section{Methods} \label{methods}

This study proposes a technique that improves ASR robustness to packet loss scenarios while maintaining the pre-trained ASR architecture and weights. As stated earlier, one option would be to use a PLC module to reconstruct the speech and subsequently apply ASR on the resulting speech. However, this solution is sub-optimal as the PLC model can introduce artifacts detrimental to the ASR model. Here, we would like to consider a different approach when replacing the PLC module with a module that will adapt the signal explicitly to improve the ASR robustness rather than make the speech sound better. We start by presenting the notation and our general setting.

We denote the speech signal by $X=(x_1, \ldots, x_T)$ as a sequence of $T$ frames (here, each $x_t$ denotes a frame of the mel-spectrum). We denote the corrupted speech signal by $\tilde{X}$, where $\tilde{x}_{k} \ldots \tilde{x}_{k+j}$ are $j$ lost frames starting at the $k$-th frame. There might be several spans of packet loss within a single utterance. We assume a transcript is associated with the speech signal, which is a sequence of $U$ words or sub-words (tokens). It will be denoted by $Y=(y_1, \ldots, y_U)$. Note that $T$ and $U$ differ for each input (and target) sequence. In our setting, we would like to propose a model that receives the corrupted speech $\tilde{X}$ and outputs the target transcription $Y$ as if it had received the original (unobserved) signal $X$.

Our model has two main components: a front-end \emph{adaptation} network, and a \emph{frozen} ASR. We denote the ASR model $\asr$ with a parameter set $\phi$. This function $\tilde{Y} =\asr(\tilde{X})$ gets as input a speech signal and predicts the word (token) sequence spoken. It is trained with some loss function $L(\asr(\tilde{X}), Y)$.

We aim to design the adaptation network $\adaptor$ with a parameter set $\theta$. This network gets as input the noisy speech and outputs an adapted version of it $\hat{X}=\adaptor(\tilde{X})$, which is used as input to the ASR, $\hat{Y}=\asr(\hat{X})$. Our goal is to have $\WER(\hat{Y},Y) \le \WER(\tilde{Y},Y)$. We note that $\hat{X}$ is generated to improve the ASR performance and might not improve human intelligibility.

This study uses Whisper models as the ASR models and a U-net architecture as the adaptation network. This network is trained with two loss functions. The first loss function is the ASR model's principal loss function, which for Whisper is cross entropy $L_\text{CE}$. This loss function guides the adapter network toward generating a spectrum better suited for greater token classification accuracy. We found that training on the ASR loss alone can sometimes converge in an unstable manner, so we added a second loss function, the $L_1$ loss component between the original signal $X$ and adapted signal $f_\theta(\tilde{X})$. This loss serves as a form of regularization, i.e., 
\begin{equation}
\min_{\theta} ~\lambda L_{\text{CE}}\big(
g_\phi(f_\theta(\tilde{X}), Y)
\big)
+ (1-\lambda) L_1\big(X, f_\theta(\tilde{X}) \big)~.
\end{equation}
We emphasize that the minimization is over the adapter network parameters $\theta$, while the Whisper $\phi$ parameters are fixed. In the evaluation, we show the advantage of our model over fine-tuning Whisper $\phi$.

\begin{figure}
    \centering
    \includegraphics[height=9cm, width=0.85\linewidth]{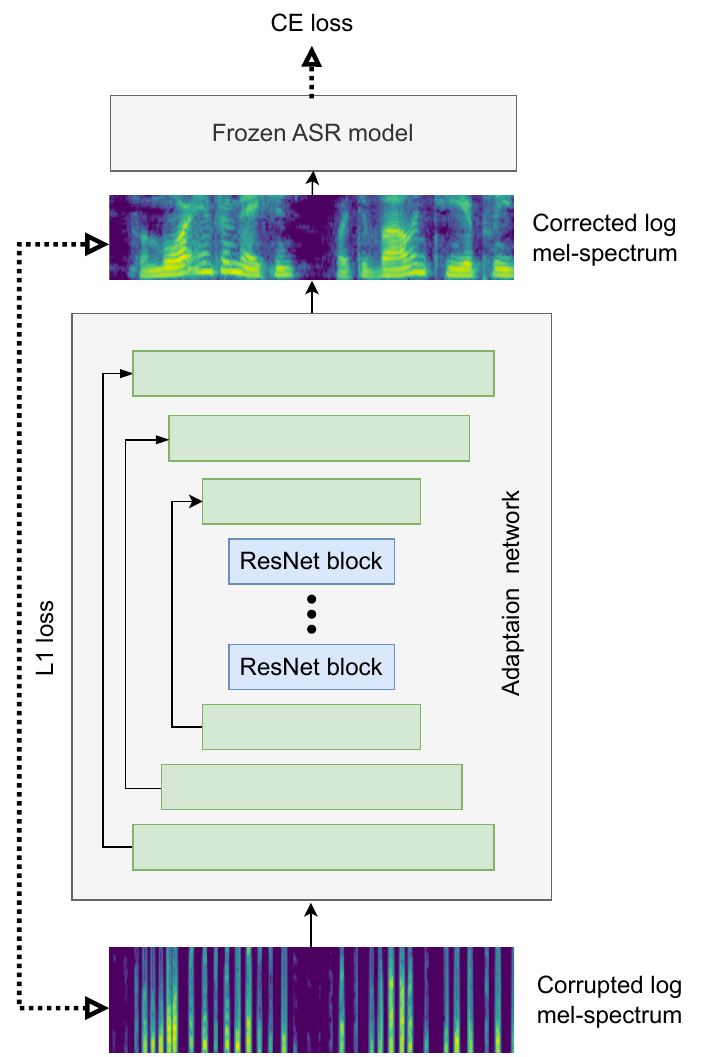}
    \caption{Our model comprises an adaptation network connected to an ASR model. The network is a convolutional U-net that receives the corrupted speech and outputs a mel-spectrum. The ASR model is a trained Whisper model.}
    \label{fig:model_arch}
\end{figure}

The adaptation network is a fully convolutional network with a U-net architecture and skip connections. The bottleneck consists of residual-blocks. Downscaling is done by maxpooling. Upscaling is done by nearest neighbor resizing followed by a convolutional layer. The input to the Whisper model is a mel-spectrum. Hence, the adapter network is designed to receive the mel-spectrum of the noisy signal $\tilde{X}$ and output an adapted mel-spectrum. This is depicted in Figure \ref{fig:model_arch}.

\begin{figure}[t]
  \centering

  \makebox[\linewidth][c]{\includegraphics[width=1.2\linewidth, trim=0 0 0 15mm, clip]{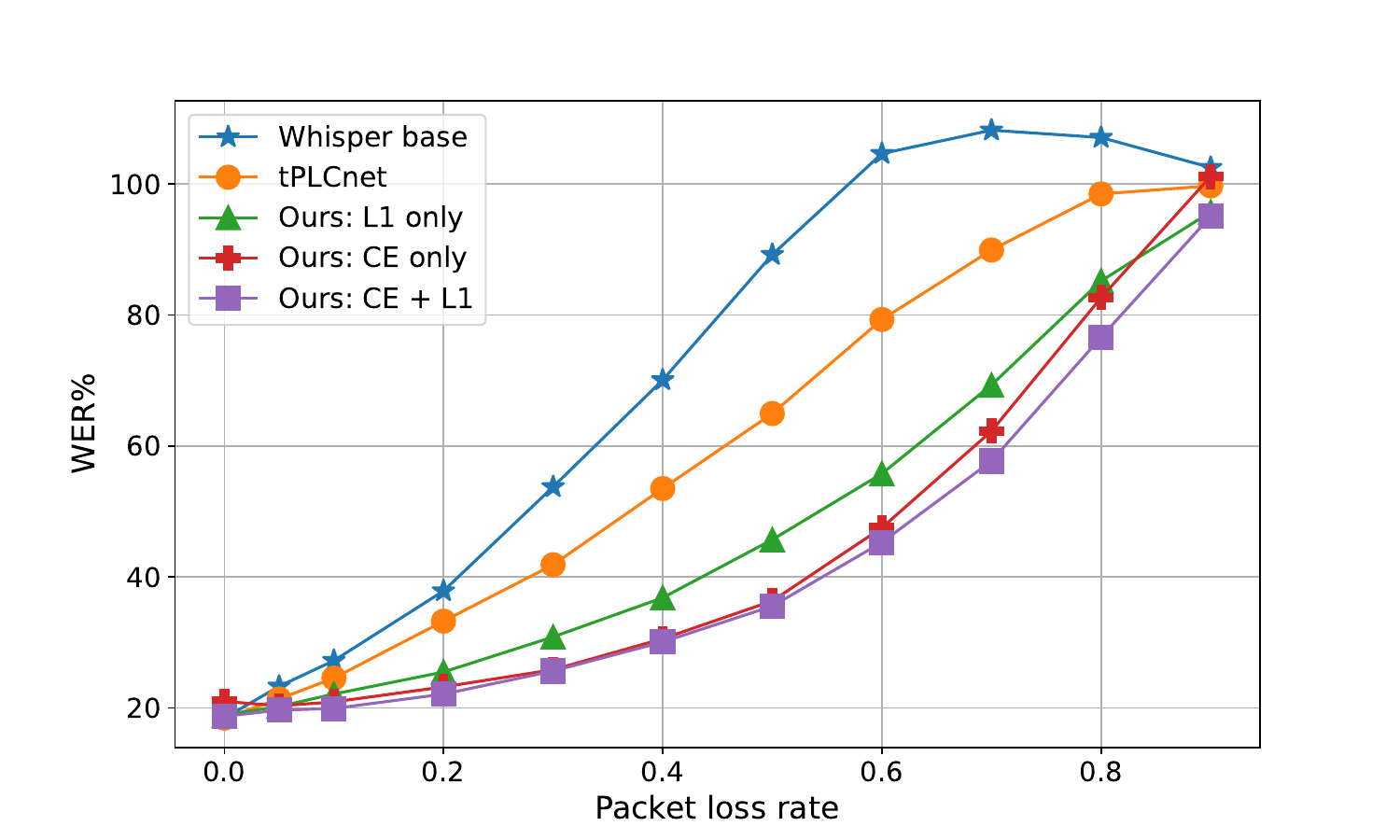}}%
  \caption{WER\% of different models on ALLSSTAR with various PLRs. All the decoding is done with Whisper base. tPLCnet refers to Westhausen and Meyer \cite{westhausen2022tplcnet}.}
  \label{fig:base}
\end{figure}
\begin{figure}[t]
  \centering
   \makebox[\linewidth][c]{\includegraphics[width=1.2\linewidth, trim=0 0 0 15mm, clip]{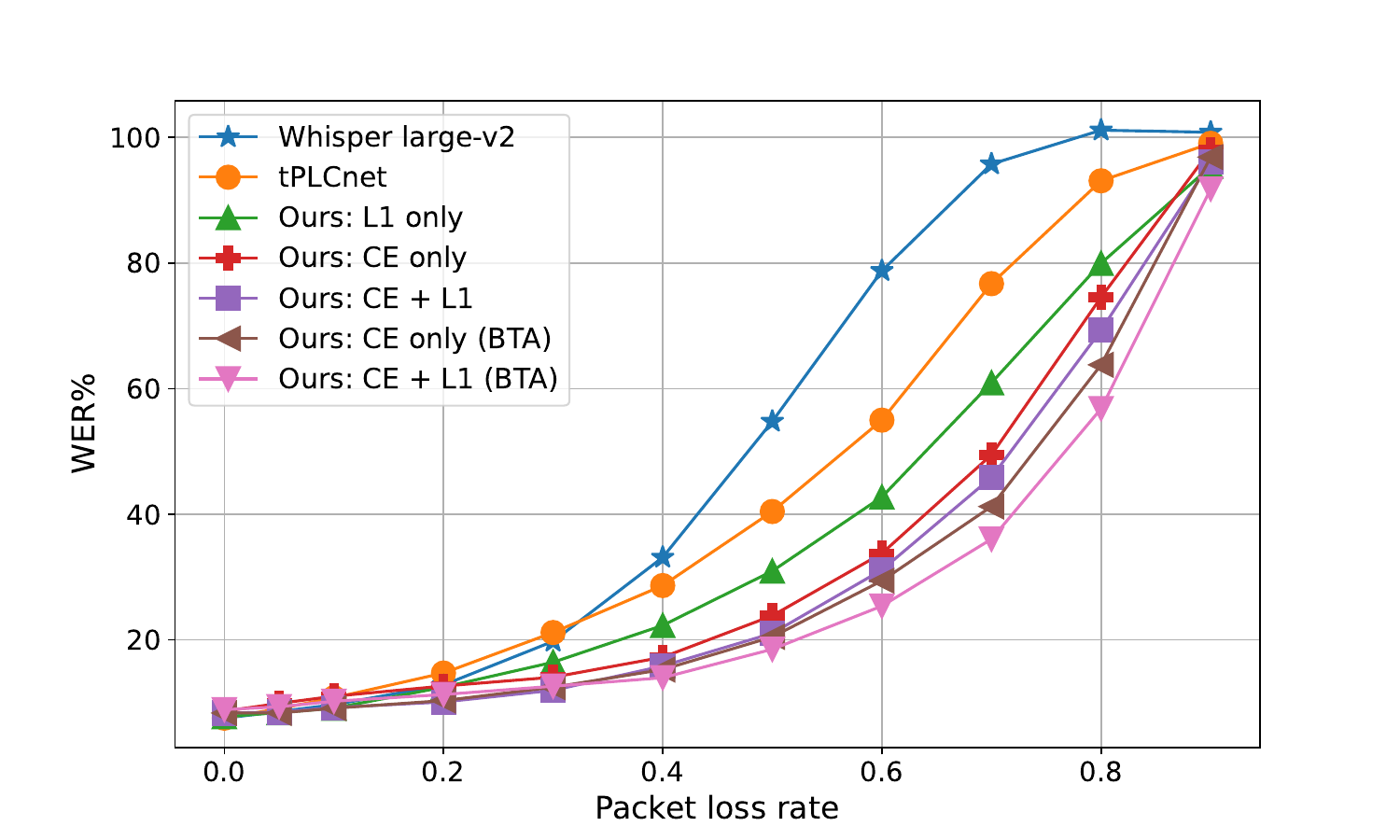}}
  \caption{WER\% of different models on ALLSSTAR with various PLRs. All the decoding is done with Whisper large-v2. tPLCnet refers to \cite{westhausen2022tplcnet}. BTA refers to adaptation networks trained on Whisper base but connected to Whisper large-v2.}
  \label{fig:large}
\end{figure}

\section{Empirical Evaluation}  \label{results}

In this section, we describe a set of experiments to demonstrate the proposed method's effectiveness empirically. We start by defining the datasets used for evaluation.

\subsection{Datasets}

For training, we use the 960 hours of English LibriSpeech \cite{panayotov2015librispeech}. For evaluation, we use multiple datasets from very different domains to showcase  the method's robustness. A subset of ALLSSTAR \cite{bradlow2010allsstar}, which is a collection of L1 Mandarin speakers speaking English \cite{kim2024automatic}, and Fleurs \cite{conneau2023fleurs} for testing on multiple languages. We don't report improvements on LibriSpeech test as they are not interesting since the improvements can come from overfitting to the training domain. 

For the packet-loss simulation, we randomly zero out frames based on two probabilities: a drop frequency (the percentage of zeroed frames per utterance) and a probabilistic distribution governing the span of consecutive frame losses. A single utterance can have multiple spans of packet loss. During training, there is a drop frequency distribution, and each sample loaded accordingly gets assigned a drop rate. During inference, for reporting reasons, we duplicate the test set to multiple fixed drop frequencies. Due to the nature of the span length distribution, there might be tiny (up to a tenth of a percent) variations from the fixed rate. When we report packet loss percentage, we mean the total percentage of lost frames in the utterance.

Additionally, we evaluated the models on the blind set from the Interspeech 2022 Audio Deep PLC Challenge \cite{diener2022interspeech} for further validation of the results. Naturally, the data for this challenge has a set packet loss, so it's used as is.  

\subsection{Experimental setting}

Recall that our model consists of two components: a frozen ASR model and an adaptation network. Whisper, which serves as the ASR model, comes in several parameter sizes. In this study, we demonstrated the effectiveness of our method with the multilingual versions of base (74M) and large-v2 (1550M) model sizes. We utilized the same Whisper parameters for all decoding, namely beam size $5$, without timestamps, and manually set the language.


The input to the adaptation network is a mel-spectrum, and the output is an estimated mel-spectrum of the same dimensions. The network is composed of three downsampling and upsampling layers, with skip connections between each equivalently sized layer, with 6 ResNet \cite{he2016deep} blocks serving as the bottleneck layers. Additionally, there are single input and single output convolutional layers that retain the same dimensions. This model's total number of trainable parameters is 7.5M, making this a negligible addition to Whisper. It is trained with the cross entropy and the $L_1$ loss functions, where the loss functions are scaled by $0.9$ for the $L_1$ and $0.1$ for the $L_\text{CE}$ (chosen on a validation set). We used a learning rate of 0.0005 with a $10\%$ decay rate per epoch. 

\subsection{Results}

In this section, we present the evaluation of the proposed method and analyze the effect of different loss functions on WER. We demonstrate the relative improvement of the proposed method over the unchanged baseline Whisper model and a recently published, open-source PLC model \cite{westhausen2022tplcnet}. We then evaluate the model's robustness to different domains and compare it to fine-tuning Whisper. In all the experiments, the Whisper baselines use \emph{zero-fill} for the dropped frames. 

Figures \ref{fig:base} and \ref{fig:large} present the performances of Whisper base and Whisper large-v2, respectively, on the original mel-spectrums in comparison with the spectrums generated by our adaptation networks and by a PLC model. The graphs present WER\% for various packet loss rate (PLR) values on the ALLSSTAR dataset. 
We note that the vanilla Whisper large model is more robust to frame loss, only starting to seriously degrade at PLRs larger than 20\%, whereas the base model starts degrading immediately. 

We present the effect of training the adaptation network with each loss function. Specifically, we compare the performance while training (i) solely using the CE loss function, $L_\text{CE}$, where the gradients flow from Whisper (noted as CE only); (ii) solely $L_1$ loss between the clean and lossy signals without referencing Whisper (noted as L1 only), which can be seen as similar to the TF-Unet in \cite{nair2021cascaded}; and (iii) a combined loss of $L_\text{CE}$ and $L_1$ (CE + L1). 

It can be seen that all methods improve WER over the original whisper model. However, the CE loss improves results more than the $L_1$ enhancement loss, whereas combining the two losses generates the most significant improvement. In Figure \ref{fig:large}, we also presented the performance of the adaptation network trained on the gradients of the base model (the one that is depicted in Figure \ref{fig:base}) but connected and evaluated with Whisper large-v2. In Figure \ref{fig:large}, we denote these models as \emph{Based Trained Adaptation (BTA)} and labeled them with \emph{Ours: CE (BTA)} and \emph{Ours: CE+L1 (BTA)}). Interestingly, training the model on Whisper base and connecting it to Whisper large-v2 gets better results than the models trained directly using Whisper large-v2. We assume this is because the gradients of the base model are easier to handle and, therefore, more effectively influence the adaptation networks. This suggests that a better training parameters exist for the large model. We defer this issue for further research. This example underscores the broader principle, that applying ASR metrics in PLC model training, can significantly enhance ASR performance across a range of models.

\begin{table}[t]
\centering
\caption{Comparison of WER\% for Different Models on the Packet Loss Challenge blind set.}
\begin{tabular}{|c|c|c|}
\hline
Model & WER\% (base) & WER\% (large-v2) \\
\hline
Whisper & 24.0 & 15.4 \\
\other ~\cite{westhausen2022tplcnet} & 20.4 & 16.2 \\
\ours & \textbf{18.1} & \textbf{14.2} \\

\hline
\end{tabular}
\label{table:PLC_challenge}
\end{table}

Furthermore, the graph shows the WER\% of tPLCnet \cite{westhausen2022tplcnet}, a time-domain many-to-one RNN model for PLC trained with a combined magnitude and complex mean absolute error loss in the time-frequency domain. We ran the large version of this model on the corrupted files and then decoded them with Whisper (base and large). The graph shows that this method improves the WER. However, the models trained using ASR metrics improve the WER more drastically.

\begin{table*}[ht]
\centering
\caption{Comparison of WER\% for Different Languages using Whisper base and large-v2. }
\begin{tabular}{|c c|c|c|c|c|c|c|c|c|}
\hline
\multicolumn{2}{|c|}{Whisper Size} & \multicolumn{4}{c|}{Base} & \multicolumn{4}{c|}{Large-V2} \\
\hline 
Packet Loss Rate & Model & French & German & Russian & Spanish & French & German & Russian & Spanish \\
\hline
\multirow{2}{*}{0\%} & 
Whisper & \textbf{24.7} & \textbf{17.2} & \textbf{20.3} & \textbf{10.3} & \textbf{7.2} & \textbf{4.6} & 6.4 & 3.7 \\
& \ours & 26.3 & 18.4 & 21.8 & 10.7 & 7.6 & 4.7 & 6.4 & 3.7 \\             
\hline

\multirow{2}{*}{5\%} & 
Whisper & 29.1 & 20.8 & 24.6 & 12.7 & \textbf{7.5} & \textbf{4.7} & \textbf{6.5} & \textbf{3.7} \\
& \ours & \textbf{27.7} & \textbf{19.6} & \textbf{23.0} & \textbf{11.2} & 7.8 & 4.8 & 6.7 & 3.9 \\ 
\hline

\multirow{2}{*}{10\%} & 
Whisper & 33.8 & 25.3 & 28.7 & 15.8 & 8.5 & 5.0 & 6.8 & 3.9 \\
& \ours & \textbf{29.3} & \textbf{20.9} & \textbf{24.7} & \textbf{11.5} & \textbf{8.1} & 5.0 & 6.8 & 3.9 \\ 
\hline

\multirow{2}{*}{20\%} & 
Whisper & 48.6 & 39.0 & 39.1 & 24.7 & 10.3 & 6.0 & 7.9 & 4.2 \\
& \ours & \textbf{32.6} & \textbf{23.7} & \textbf{27.9} & \textbf{13.2} & \textbf{8.8} & \textbf{5.7} & \textbf{7.5} & \textbf{4.1} \\  
\hline
\multirow{2}{*}{30\%} & 
Whisper & 69.5 & 60.7 & 53.7 & 38.3 & 16.0 & 8.3 & 11.5 & 5.3 \\
& \ours & \textbf{38.3} & \textbf{27.2} & \textbf{32.6} & \textbf{15.5} & \textbf{9.9} & \textbf{6.3} & \textbf{8.8} & \textbf{4.4} \\  
\hline
\multirow{2}{*}{40\%} & 
Whisper & 104.9 & 102.2 & 75.6 & 57.2 & 27.5 & 13.4 & 18.2 & 7.5 \\
& \ours & \textbf{41.5} & \textbf{32.8} & \textbf{39.0} & \textbf{19.0} & \textbf{12.3} & \textbf{7.6} & \textbf{9.9} & \textbf{5.2} \\  
\hline
\multirow{2}{*}{50\%} & 
Whisper & 126.7 & 141.2 & 100.8 & 89.2 & 48.3 & 26.4 & 34.4 & 12.7 \\
& \ours & \textbf{49.6} & \textbf{40.4} & \textbf{46.1} & \textbf{24.4} & \textbf{15.5} & \textbf{10.3} & \textbf{12.7} & \textbf{6.2} \\  
\hline
\multirow{2}{*}{60\%} & 
Whisper & 124.6 & 140.4 & 121.1 & 127.9 & 71.7 & 51.8 & 61.5 & 26.6 \\
& \ours & \textbf{61.2} & \textbf{53.2} & \textbf{59.2} & \textbf{33.7} & \textbf{20.6} & \textbf{15.5} & \textbf{19.1} & \textbf{9.4} \\  
 
\hline
\end{tabular}
\label{table:multilingual}
\end{table*}

Next, in Table \ref{table:PLC_challenge}, we compare the WER of the baseline whisper models, tPLCnet \cite{westhausen2022tplcnet} large and \ours{} on the blind set from the interspeech 2022 PLC challenge \cite{diener2022interspeech}. Here, the PLRs are set by the challenge. As seen, tPLCnet improves the WER over the baseline model but not Whisper large, and \ours{} performs best in both scenarios. 

To further demonstrate the model's robustness to domains and to showcase that this training method doesn't harm the original Whisper models as opposed to fine-tuning where training in one domain or language degrades the model's performance in other domains or languages, we compare, in Table \ref{table:multilingual}, the WER of the model to the original Whisper models in multiple languages selected at random from the Fleurs dataset \cite{conneau2023fleurs}. Here, the pattern is similar to the results on the ALLSSTAR dataset, as shown in Figures \ref{fig:base} and \ref{fig:large}: the base model starts degrading immediately, and the large only after 20\% PLR, whereas \ours, other than a slight degradation in the zero PL scenario, improves results for all PLRs in all languages.

\begin{table}[t]
\centering
\caption{Comparison of WER\% for finetuning vs \ours{} using Whisper base.}
\begin{tabular}{|l|c c c|c c c|}
\hline
Dataset &
\multicolumn{3}{|c|}{ALLSSTAR} & \multicolumn{3}{c|}{Spanish} \\
\hline
{PLR} & 0 & 0.2 & 0.4 & 0 & 0.2 & 0.4 \\
\hline
Whisper   & \textbf{18.4}\hspace{-0.1cm} & 37.8\hspace{-0.1cm} & 70.0\hspace{-0.1cm} & \textbf{10.3}\hspace{-0.1cm} & 24.7\hspace{-0.1cm} & 57.2\hspace{-0.1cm} \\
Fine-tune & 24.9\hspace{-0.1cm} & 27.1\hspace{-0.1cm} & 31.8\hspace{-0.1cm} & 89.5\hspace{-0.1cm} & 91.1\hspace{-0.1cm} & 94.5\hspace{-0.1cm} \\
\ours     & 18.7\hspace{-0.1cm} & \textbf{22.1}\hspace{-0.1cm} & \textbf{30.0}\hspace{-0.1cm} & 10.7\hspace{-0.1cm} & \textbf{13.2}\hspace{-0.1cm} & \textbf{19.0}\hspace{-0.1cm} \\

\hline
\end{tabular}
\label{table:finetune}
\end{table}

Finally, to establish the advantage of this method over fine-tuning, we also fine-tuned Whisper using the same training data, and the outcomes aligned precisely with our earlier theories. Notably, the performance on the LibriSpeech test set improved substantially. However, this enhancement came at a cost: first, as seen in Table \ref{table:finetune}, the model lost its multilingual capabilities; second, its performance on ALLSSTAR, an English dataset but from a different domain, experienced a significant decline from the baseline in the no packet loss (clean) case.

\section{Related work}  \label{related}

\noindent {\bf PLC.} While packet loss concealment is well studied, it primarily focuses on perceptual audio quality, focusing on improving metrics like Perceptual Evaluation of Speech Quality (PESQ) and Short-Time Objective Intelligibility (STOI), whereas this research explicitly targets ASR performance. The current more prevalent framework for neural network-based PLC are encoder decoder speech inpainting methods such as Wang \emph{et al.}, \cite{wang2021temporal} or Pascual \emph{et al.}, \cite{pascual2021adversarial} who use adversarial training to generate the most natural sounding audio in the gaps. Another standard method is a causal, recurrent, sequence-to-one model where the objective is to predict the next frame given the previous frames. Westhausen and Meyer \cite{westhausen2022tplcnet} whom we compared to, predict the next lost frame with a short context buffer in the time domain, using an RNN. Lin \emph{et al.} \cite{lin2021time} frame it as a generative regression problem and utilize a convolutional encoder-decoder with LSTM layers for next frame prediction in the time domain.

\noindent {\bf Spectral inpainting.} Context-based retrieval of missing parts of time-frequency representation of speech using a convolutional U-net was originally demonstrated by Kegler \emph{et al.}\cite{kegler2019deep}. Simon \emph{et al.} \cite{simon2022correcting}, used it for correcting mispronunciations in speech by cutting out the mispronunciations and inpainting it correctly  with the context.

\noindent {\bf ASR guided enhancement. }
There is also precedent for using ASR objectives to improve distorted signals. Subramanian \emph{et al.} \cite{subramanian2019speech} used end-to-end speech recognition objectives to train a speech enhancement model. They showed that in addition to improving the WER, they also improved the speech enhancement metrics. Another interesting method, presented by Yang \emph{et al.} \cite{yang2023towards}, uses an auxiliary loss between latent ASR representations of the clean signal and the representations generated by their packet loss concealment model. This generates more natural speech reconstruction.

\section{Discussion and future work} \label{future}

This study introduced a novel approach to enhance the reliability of large ASR models in facing packet loss scenarios. The proposed method involves the integration of a smaller model, which is specifically designed to adapt the input features of an ASR model. The study has demonstrated that this integration process leads to significant improvements in the robustness of the ASR models when trained using the gradients of a larger ASR model.
The promising results shown by the proposed method opens up new avenues for future research in the field of ASR development.
Future research can investigate the applicability of this method in enhancing the robustness of ASR models against various noise types, such as white noise, babble, clipping, and echo suppression. It would also be interesting to examine the generalizability of this approach across different ASR model architectures, such as HuBERT, and evaluate the extent to which this approach can improve traditional intelligibility metrics. 

\bibliographystyle{IEEEtran}
\bibliography{mybib}

\end{document}